\documentclass[prd,preprint,superscriptaddress,preprintnumbers,nofootinbib]{revtex4}
\usepackage{graphicx}
\usepackage{epsfig}
\usepackage{bm}
\usepackage{latexsym,amssymb,amsmath,amssymb,wasysym,float}
\usepackage{mathrsfs}  
\usepackage{color}

\usepackage{enumitem}

\newcommand{\postscript}[2]{\setlength{\epsfxsize}{#2\hsize}
   \centerline{\epsfbox{#1}}}

\usepackage[usenames,dvipsnames]{xcolor}
\definecolor{orange}{cmyk}{0,0.5,1,0}
\definecolor{rossoCP3}{cmyk}{0,.88,.77,.40}
\definecolor{graa}{rgb}{0.8,0.8,0.8}
\definecolor{blaa}{rgb}{0.2,0.2,0.6}

\begin{document}

\preprint{MPP-2022-285}
\preprint{LMU-ASC 55/22}

\title{\color{rossoCP3} Aspects of the Dark Dimension in Cosmology}

\author{\bf Luis A. Anchordoqui}

\affiliation{Department of Physics and Astronomy,\\  Lehman College, City University of
  New York, NY 10468, USA
}

\affiliation{
 Graduate Center, City University
  of New York,  NY 10016, USA
}

\affiliation{Department of Astrophysics,
 American Museum of Natural History, NY
 10024, USA
}

\author{\bf Ignatios Antoniadis}

\affiliation{Department of Physics, Harvard University, Cambridge, MA 02138, USA}

\affiliation{Laboratoire de Physique Th\'eorique et Hautes \'Energies - LPTHE,\\
Sorbonne Universit\'e, CNRS, 4 Place Jussieu, 75005 Paris, France
}

\author{\bf Dieter\nolinebreak~L\"ust}

\affiliation{Max--Planck--Institut f\"ur Physik,  
 Werner--Heisenberg--Institut,
80805 M\"unchen, Germany
}

\affiliation{Arnold Sommerfeld Center for Theoretical Physics, \\
Ludwig-Maximilians-Universit\"at M\"unchen,
80333 M\"unchen, Germany
}

\begin{abstract}
  \noindent It was recently understood that if the swampland
  conjectures are confronted to experiment they naturally point to a
  solution of the cosmological hierarchy problem in which the smallness of
  the dark energy is ascribed to an internal (dark) dimension with characteristic
length-scale in the micron range. It was later inferred that
the universal coupling of the Standard Model fields to the massive
spin-2 Kaluza-Klein (KK) excitations of the graviton in the dark
dimension leads to a dark matter candidate. Since the
  partial decay widths of KK gravitons into the visible sector must be
  relatively small to accommodate experiment, the model is
  particularly challenging to probe.  We show that the model can accommodate
  neutrino masses associated to right-handed neutrinos propagating in
  the bulk of the dark dimension with an additional constraint imposed
  by neutrino oscillation data. After that, we study the impact of the KK
  tower in cosmology. We show that the modulation of redshifted 21-cm
  lines driven by ${\rm KK} \to \gamma \gamma$ could be within the
  reach of next generation experiments (e.g. SKA and
  FARSIDE). We also show
  that indirect dark matter searches could uncover the  ${\rm KK}
  \to \gamma \gamma$ signal. These two
  observations combined have the potential for model identification. Finally, we explore
  the global structure of the inflationary phase and demonstrate that
the model parameters required for a successful uniform inflation driven by a 5-dimensional
  cosmological constant  (corresponding to a flat region of the 5-dimensional potential) are natural. 
\end{abstract}

\maketitle

\section{Introduction}

It was recently pointed out that by combining the distance
conjectures~\cite{Ooguri:2006in,Lust:2019zwm} of the Swampland
program~\cite{Vafa:2005ui} with the smallness of the dark energy in
Planck units ($\Lambda \sim 10^{-120} M_{\rm Pl}^4$) and confronting
these ideas to experiment~\cite{Kapner:2006si,Lee:2020zjt,Hannestad:2003yd} lead to
the prediction of a compact dark dimension with characteristic
length-scale in the micron range~\cite{Montero:2022prj}. The dark
dimension opens up at the characteristic mass scale of the
Kaluza-Klein (KK) tower, $m_{\rm KK} \sim \Lambda^{1/4}/\lambda$, where
physics must be described by a 5-dimensional theory up to the ``species
scale'', $M_{\rm UV} \sim \lambda^{-1/3} \ \Lambda^{1/12}
\ M_{\rm Pl}^{2/3}$, which can be regarded as the higher dimensional Planck
scale, with $10^{-1} \alt \lambda \alt 10^{-4}$.\footnote{Throughout
  we defined the species scale in terms of the reduced Planck mass
  $M_{\rm Pl}$ rather than the Planck mass as in~\cite{Dvali:2007hz,Dvali:2007wp}.} The dark
dimension model carries with it a rich
phenomenology~\cite{Anchordoqui:2022ejw,Anchordoqui:2022txe,Blumenhagen:2022zzw, Gonzalo:2022jac,Anchordoqui:2022tgp}. In
particular, it was observed in~\cite{Gonzalo:2022jac} that
the universal coupling of the Standard Model (SM) fields to the massive
spin-2 KK excitations of the graviton in the dark dimension provides a dark matter candidate. Complementary to the dark gravitons, it was discussed in~\cite{Anchordoqui:2022txe} that primordial black holes with Schwarzschild radius smaller
than a micron could also be good dark matter candidates, possibly even with an interesting close relation to the dark gravitons~\cite{Anchordoqui:2022tgp}.

In the dark dimension graviton (DDG)
model the cosmic evolution of the hidden sector is primarily dominated by ``dark-to-dark'' decays, yielding
  a specific realization of the dynamical dark matter
  framework~\cite{Dienes:2011ja}. The tower of KK states propagating
  in the incredible bulk provide a rich playground for novel
  cosmological signals. Herein, we confront DDG predictions to experiment and demonstrate it may represent a viable alternative to the $\Lambda$ cold-dark-matter (CDM) model.

  The layout of the paper is as follows.  We begin in
  Sec.~\ref{sec:nu} with a concise summary of neutrino masses, mixing,
  and oscillations within the context of bulk-neutrino
  models~\cite{Dienes:1998sb,Arkani-Hamed:1998wuz,Dvali:1999cn,Davoudiasl:2002fq,Antoniadis:2002qm}.\footnote{Neutrino
    masses in relation to the swampland program were discussed
    in~\cite{Ibanez:2017kvh,Gonzalo:2021zsp}.}  Using observational
  bounds on large extra dimensions we obtain constraints on the
  parameter space of the DDG model enriched with bulk right-handed
  neutrinos. In Sec.~\ref{sec:2} we develop a two-step test to
  unambiguously distinguish DDG predictions from those of
  $\Lambda$CDM and its extensions.  High-redshift post-reionization
  cosmology with 21-cm lines allow us to perform a tomographic study
  of the dark ages and cosmic dawn. We show that there is a promising
  region of the DDG parameter space in which the modulation of
  redshifted 21-cm lines driven by ${\rm KK} \to \gamma \gamma$ will 
  be within the reach of next generation experiments. Exploring in
  the same promising region of the parameter space, we estimate the
  evolution of dark matter over the age of the universe to pin down
  the mass of the dark graviton today. Choosing DDG parameters from
  the indicated domain, we demonstrate that the sensitivity of
  indirect dark matter searches targeting dwarf spheroidal galaxies
  are extremely competitive to detect the KK decay products in the
  local universe. Due to the particular evolution of the dark matter 
  mass, a successful observation of both  low- and high-redshift
  signals would allow a well-defined identification of the DDG model. In Sec.~\ref{sec:3} we explore
  the global structure of the inflationary phase within the context of
  the dark dimension. We demonstrate that
the model parameters required for a successful uniform inflation driven by a 5-dimensional
  cosmological constant  (corresponding to a flat region of the
  5-dimensional potential) are natural. The paper wraps up with some
  conclusions presented in Sec.~\ref{sec:4}.

  \section{Constraints from neutrino oscillation experiments}
\label{sec:nu}

We adopt the working assumption that neutrino masses originate in 
three 5-dimensional fermion fields $\Psi_\alpha \equiv
(\psi_{\alpha L},\psi_{\alpha R})$, which 
are singlets under the SM gauge symmetries and interact on our brane with the three active left-handed neutrinos
  $\nu_{\alpha L}$ in a way that conserves lepton number, where the indices $\alpha = e, \mu,\tau$ denote the
  generation~\cite{Dienes:1998sb,Arkani-Hamed:1998wuz,Dvali:1999cn,Davoudiasl:2002fq,Antoniadis:2002qm}. From a 4-dimensional perspective, each of the singlet fermion fields can be decomposed as
  an infinite tower of KK states, $\psi^\kappa_{L(R)}$,
  with $\kappa = 0,\pm 1, \cdots, \pm \infty$. The right-handed bulk
  states $\psi_R^\kappa$ combine with the left-handed bulk components
  $\psi_L^\kappa$ to form Dirac
mass terms, which derive from the quantized internal momenta in the
dark dimension. The bulk states also
mix with the active
left-handed neutrinos through Dirac-like mass terms. Redefining the bulk fields as $\nu^{(0)}_{\alpha R} \equiv\psi^{(0)}_{\alpha R}$ and
$\nu^{(K)}_{\alpha L(R)}\equiv \Big(\psi^{(K)}_{\alpha
  L(R)}+\psi^{(-K)}_{\alpha L(R)}\Big)/\sqrt 2$, after electroweak
symmetry breaking the
mass terms of the Lagrangian take the form
\begin{eqnarray}
\mathscr{L_{\text{mass}}}  &= & \displaystyle
\sum_{\alpha,\beta}m_{\alpha\beta}^{D}\left[\overline{\nu}_{\alpha
    L}^{\left(0\right)}\,\nu_{\beta R}^{\left(0\right)}+\sqrt{2}\,
  \sum_{K=1}^{\infty}\overline{\nu}_{\alpha
    L}^{\left(0\right)}\,\nu_{\beta
    R}^{\left(K\right)}\right]  
 +  \sum_{\alpha}\sum_{K=1}^{\infty}\displaystyle
m_K \, \overline{\nu}_{\alpha L}^{\left(K\right)} \,
\nu_{\alpha R}^{\left(K\right)} 
                                \, + {\rm h.c.} \nonumber \\
& = & \sum_{i=1}^3 \bar{\mathbb{N}}_{iR} \  \mathbb
      M_i \ \mathbb{N}_{iL}+ {\rm h.c.}  \,,
      \label{calL}
\end{eqnarray}
where $m_{\alpha \beta}^{D}$ is a Dirac mass
matrix, $m_K = K/R = K m_{\rm KK}$,
\begin{eqnarray}
{\mathbb{N}_{i L(R)}}=\Big(\nu_i^{(0)},\nu_i^{(1)},\nu_i^{(2)},\cdots\Big)^T_{L(R)},
  ~~~~~{\rm{and}}~~~~~ \mathbb M_i=
\begin{pmatrix}
m_i^D&0&0&0&\ldots\\
\sqrt{2}m_i^D&1/R&0&0&\ldots\\
\sqrt{2}m_i^D&0&2/R&0&\ldots\\
\vdots&\vdots&\vdots&\vdots&\ddots
\end{pmatrix},
\end{eqnarray}
and where $m_i^D$ are the elements of the diagonalized Dirac mass
matrix $=\mathrm{diag}(m^D_1,m^D_2,m^D_3)$. Greek indices from the
beginning of the alphabet run over the 3 active flavors ($\alpha,\beta
= e,\mu,\tau$), Roman lower case indices over the 3 SM families $(i =
1,2,3)$, and capital 
Roman indices over the KK modes ($K =1,2,3,..., +\infty$). Note that
$\psi_{\alpha L}^{(0)}$ decouples from the system. For the
configuration at hand,
\begin{equation}
m_i^D \approx \frac{y_i \ \langle H\rangle}{\sqrt{R M_{\rm UV}}} \,,
\label{in3}
\end{equation}
where $y_i$ are the Yukawa couplings and $\langle H \rangle$ the Higgs
vacuum expectation value. 

A quantity of pivotal interest is $P(\nu_\alpha \to \nu_\beta)$, which
  defines the probability of finding a neutrino of flavor $\beta$ in a
  beam that was born with flavor $\alpha$ and has travelled a distance
  $L$. Armed with the Lagrangian (\ref{calL}) it is straightforward to
  parametrize $P(\nu_\alpha \to \nu_\beta)$ in terms of three mixing angles
    ($\theta_{12}, \theta_{13}, \theta_{23}$), a Dirac CP-violating
    phase ($\delta_{13}$), and the solar ($\Delta m^2_{21}$) and
    atmospheric ($\Delta m^2_{31}$) mass
    differences~\cite{Dienes:1998sb,Arkani-Hamed:1998wuz,Dvali:1999cn,Davoudiasl:2002fq,Antoniadis:2002qm}. Short- 
    and long-baseline experiments constrain the standard 6 oscillation
    parameters and two extra parameters taken to be $R$ and $m_0 \equiv m_{1(3)}^D$
    for the normal hierarchy $m_3>m_2 > m_1 = m_0$ and (inverted
    hierarchy $m^D_2>m^D_1> m^D_3 =
    m_0$)~\cite{Davoudiasl:2002fq,Machado:2011jt}. The most recent analysis of neutrino oscillations, combining data from MINOS/MINOS+, Daya Bay, and
    KATRIN, gives $R < 0.4~\mu {\rm m}$, for the normal
    hierarchy, and $R < 0.2~\mu{\rm m}$, for the inverted
    hierarchy, both upper limits at
    99\%CL~\cite{Forero:2022skg}.\footnote{A point
      worth noting at this juncture is that in the presence of bulk
      masses, right-handed neutrinos propagating in the dark dimension can
      induce electron-neutrino appearance effects~\cite{Carena:2017qhd}. These effects can
      help relax the
      bound on $R$ and have the potential to address the LSND $\overline
      \nu_\mu \to \overline \nu_e$ anomaly~\cite{LSND:2001aii}.} This implies
    that $m_{\rm KK} \agt 2.5~{\rm eV}$, and therefore
    $\lambda \alt 10^{-3}$. The associated upper limit on the
    compactification radius $R$ is two orders of magnitude below the
    previously predicted maximum size of the dark dimension. The less
    restrictive bound is based on
    null results in the search for deviations from Newton's
    gravitational inverse-square law in the short length-scale regime~\cite{Kapner:2006si,Lee:2020zjt}.

\section{Two-Step Test of Dynamical Dark Graviton Cosmology}
\label{sec:2}

This section describes an overall framework for DDG model
identification. We show that observations at low- and high-redshift
from next generation experiments could be used to unequivocally identify the
cosmic evolution of the incredible bulk~\cite{Dienes:2011ja}. As an
illustration, we proceed here with normalization of the model
parameters at high redshift to establish model predictions for the
local universe; but of course, this direction can be turned the other way around.

\subsection{Probing the Dark Dimension with Highly-Redshifted 21-cm Line}

During the emission of the cosmic microwave background (CMB) at redshift $z_{\rm CMB} \sim 1100$, matter dominates the energy
density of the Universe. The number density of baryons is mostly
composed of neutral hydrogen atoms (${\rm H}_{\rm I}$), together with a smaller Helium
(He) component, $x_{\rm He} = n_{\rm He}/n_{{\rm H}_{\rm I}} \simeq 1/13$, and a
small percentage of free protons and electrons, $x_e = n_e /n_{{\rm
    H}_{\rm I}} =
n_p /n_{{\rm H}_{\rm I}}$, which varies from about 20\% at $z_{\rm CMB}$ to roughly
$2 \times 10^{-4}$ at $z \sim 20$~\cite{Ali-Haimoud:2010tlj}. After the gas thermally decouples from the photon
temperature at $z_{\rm dec} \sim 150$, most of the hydrogen gas is
in its ground state, whose degeneracy is only broken by the hyperfine
splitting of the spin-0 singlet and
spin-1 triplet. In the rest 
frame of the gas, the energy gap between these two spin states is $E_{21} = 5.9 \times 10^{-6} {\rm eV} \simeq
2\pi/21~{\rm cm}$, corresponding to a 
 $\nu_{21} = 1.4~{\rm GHz}$ spectral line. The relative number density of the two spin levels,
$n_1/n_0 = 3 e^{-E_{21}/T_s}$, defines the spin temperature $T_s$,
where the factor of 3 comes from the degeneracy of the triplet excited
state. Because
 of the 21-cm transitions, for a
 given redshift $z < z_{\rm dec}$, a shell of $H_{\rm I}$  can act as a
 detector of the background photons~\cite{Pritchard:2011xb}.
The observable is the 21-cm brightness temperature relative to the
photon background
\begin{equation}
T_{21} (z) \simeq 27~{\rm mK} \ x_{{\rm H}_{\rm I}} (z) \
\left(\frac{0.15}{\Omega_m h^2} \right)^{1/2} \ \left(\frac{1+z}{10}\right)^{1/2} \
\left(\frac{\Omega_b h^2}{0.02} \right)
\ \left(1 - \frac{T_\gamma(z)}{T_s (z)} \right) \,,
\label{T21}
\end{equation}  
where $\Omega_m$ and $\Omega_b$ are the the present day values of the
non-relativistic matter density and baryon energy
density as a fraction of the critical density, and where $h$ is the Hubble constant constant in units
of $100~{\rm km/s/Mpc}$~\cite{Zaldarriaga:2003du}. Note that if the
photon temperature exceeds the spin temperature ($T_{21} < 0$) there
will be net absorption and in the opposite case ($T_{21} > 0$) net
emission. For $\Lambda$CDM, $T_\gamma$ is given by the CMB thermal
radiation, with temperature $T_{\rm CMB}(z) = 2.725~{\rm K} \ (1+z)$.

From the Dark Ages ($1100 \alt z \alt 30$) to Cosmic Dawn ($30 \alt z
\alt 15$) and the subsequent Epoch of Reionization, the evolution of $T_{21}$ can be
schematically described by five distinctive regimes: {\it (i)}~For $z_{\rm
  dec} < z < z_{\rm CMB}$, the non negligible amount of free electrons
couples the gas to
radiation, $T_\gamma = T_{\rm gas} = T_s$, and there is
no 21-cm signal because $T_{21} = 0$. For $z< z_{\rm dec}$, the gas cools down
more rapidly than CMB radiation, but
 the gas temperature has a minimum value set by its primordial
 heating due to Thomson scattering and so $T_{\rm gas} (z) = T_{\rm
   CMB} (z) (1 + z)/(1 + z_{\rm
   dec})$. {\it (ii)}~For $30 \alt z < z_{\rm dec}$,
gas collisions are efficient enough to couple the spin and gas
temperatures, i.e. $T_s \simeq T_{\rm gas}$. This means $T_{\rm 21} <
0$ and therefore an absorption signal is expected during the Dark
Ages. {\it (iii)}~At $z \simeq 30$ the gas becomes so rarified that the collision rate
becomes too low to enforce $T_s \simeq T_{\rm gas}$ and the 21-cm
signal is again suppressed $T_{\rm 21} \simeq 0$ because $T_s \simeq
T_\gamma$. {\it (iv)}~At Cosmic Dawn $z \alt 30$, the first stars
fired up, making the Lyman-$\alpha$ coupling strong and thereby the
spin temperature is expected to be coupled to the gas kinetic
temperature, i.e., $T_{\rm gas} \approx T_s$. Again $T_{21} < 0$,
implying an absorption signal. {\it (v)}~During the epoch of
reionization the gas gets reheated by astrophysical radiation yielding
$T_s \simeq T_{\rm gas} > T_\gamma$, so that $T_{21}$ turns positive
and one has an emission signal from the regions that are not fully
ionized. Eventually all gas gets ionized until the fraction of neutral
hydrogen vanishes and the signal switches off again.

The Experiment to Detect the Global Epoch of Reionization Signature
(EDGES) recorded the first measurement of the global 21-cm
spectrum. The data are consistent with an absorption profile at $z
\sim 15 - 20$, with a minimum at $z_E =17.2$ where $T_{21} (z_E ) = -500^{+200}_{-500}~{\rm mK}$ at 99\% C.L., including estimates of
systematic uncertainties~\cite{Bowman:2018yin}. Using (\ref{T21}) with $x_{H_I}
\simeq 1$ the EDGES observation implies
$T_{\rm gas} \simeq T_s (z_E) < 3.3~{\rm K}$, wherefore in tension with
$\Lambda$CDM that predicts
the minimum gas temperature to be 
$T_{\rm gas}^{\Lambda{\rm CDM}} (z_E) \simeq 6~{\rm K}$.\footnote{We
  note in passing that this interpretation has been called into
  question in~\cite{Hills:2018vyr} and it is in tension with the
  SARAS3 data~\cite{Bevins:2022ajf}.}

The EDGES 21-cm signal severely constrains new physics processes which
are capable of heating up the intergalactic medium prior to the
reionization time, e.g., via energy injection from
annihilation~\cite{DAmico:2018sxd} or
decaying~\cite{Poulin:2016anj,Clark:2018ghm,Liu:2018uzy} dark matter. For sub-MeV
dark matter, EDGES bounds on the partial decay width into two photons
are stronger than CMB-based limits~\cite{Slatyer:2016qyl} by one to
two orders of magnitude.

We now turn to investigate how the EDGES
bound impacts the allowed parameter space of the DDG model. We begin by considering a tower of equally spaced dark gravitons,
indexed by an integer $l$, and with mass  $m_l = l \ m_{\rm KK}$. The partial decay width of KK graviton $l$ to SM fields is found to be,
\begin{equation}
  \Gamma^l_{{\rm SM}} = \frac{\tilde{\lambda}^2 \ m^3_{\rm KK} \ l^3}{80 \pi
    M_{\rm Pl}^2} \,,
\label{Ggamma}
\end{equation}
where $\tilde \lambda$ takes into account all the available decay channels and is a function of
time~\cite{Hall:1999mk}.

The cosmic evolution of the dark sector is mostly driven by ``dark-to-dark''
decay processes, which regulate the decay of KK gravitons within the dark tower~\cite{Gonzalo:2022jac}. In the absence of isometries
in the dark dimension, which is the common expectation, the KK momentum of the
dark tower is not conserved. This means that a dark graviton of KK
quantum $n$ can decay to two other ones, with quantum numbers $n_1$
and $n_2$. If the KK quantum violation can go
up to $\delta n$, the number of available channels is roughly $l \,
\delta n$. In addition, because the decay is almost at threshold, the phase space
factor is roughly the velocity of decay products,
  $v_{\rm r.m.s.} \sim \sqrt{m_{\rm KK} \ \delta n /m_l}$. Putting all
  this together we obtain the total
  decay width, 
\begin{equation}
  \Gamma^l_{\rm tot} \sim \sum_{l'<l} \ \ \sum_{0<l''<l-l'} \Gamma^l_{l' l''} \sim
  \beta^2 \frac{m_l^3}{M_{\rm Pl}^2} \times \frac{m_l}{m_{\rm KK}} \ \delta
  n \times \sqrt{\frac{m_{\rm KK} \delta_n}{m_l} }\sim \beta^2 \
    \delta n^{3/2} \frac{m_l^{7/2}}{M_{\rm Pl}^2 m_{\rm KK}^{1/2}} \,,
 \end{equation}   
where $\beta$ parametrizes our ignorance of decays in the dark dimension~\cite{Gonzalo:2022jac}.

We further follow~\cite{Gonzalo:2022jac} to estimate the time
evolution of the dark matter mass and
assume that for times larger than $1/\Gamma^l_{\rm tot}$ dark matter
which is
heavier than the corresponding $m_l$ has already decayed, yielding
\begin{equation}
  m_l \sim \left(\frac{M_{\rm Pl}^4 \ m_{\rm KK}}{\beta^4 \ \delta n^3}\right)^{1/7} t^{-2/7} \,,
\label{mt}
\end{equation}
where $t$ indicates the time elapsed since the big bang.

\begin{figure}[tpb]
\postscript{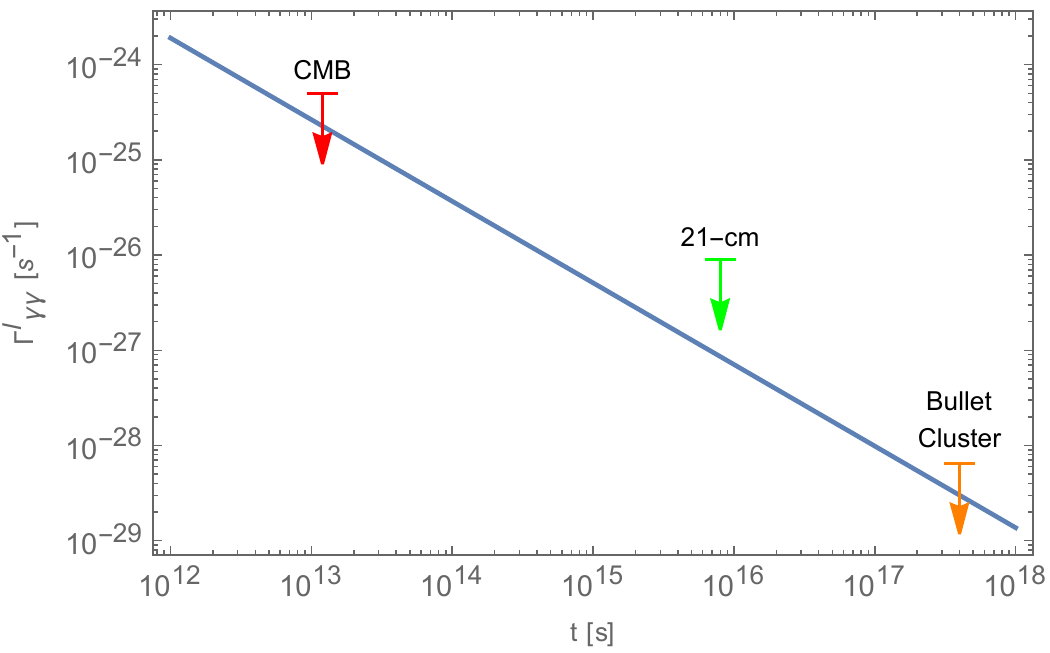}{0.9}
\caption{$\Gamma^l_{\gamma \gamma}$ as
  a function of the age of the Universe. The curve is normalized to a
  dark matter mass of 1 MeV at the time of matter-radiation
  equality. Bounds on the partial decay width of dark matter decaying
  into two photons from CMB
  anisotropies~\cite{Slatyer:2016qyl},  21-cm absorption line detected by EDGES~\cite{Clark:2018ghm,Liu:2018uzy}, and the bullet cluster~\cite{Riemer-Sorensen:2015kqa} are shown
  for comparison. The age of the Universe is taken to be $13~{\rm Gyr}$~\cite{Planck:2018vyg}. \label{fig:1}}
\end{figure}

All in all, the recombination history of the Universe would be modified
by dark gravitons decaying into SM fields, as these visible fields inject energy into the (pre-recombination) photon-baryon plasma
and (post-recombination) gas and background
radiation. The energy
injection would increase the ionization of the gas, the atomic
excitation of the gas, and the plasma/gas heating. These three processes would therefore increase the residual ionization fraction
($x_e$) and baryon temperature after
recombination. Consistency with CMB anisotropies requires  $\Gamma^l_{\gamma \gamma} < 5 \times 
10^{-25}~{\rm s}^{-1}$ between the last scattering surface and
reionization~\cite{Slatyer:2016qyl}. In our calculations we set 
$\lambda \sim 10^{-3}$ to accommodate neutrino masses with 3
generations of massless
bulk fermions. Taking $\tilde{\lambda} = 1$ to
set out the decay into photons we can use (\ref{Ggamma}) to find that the
CMB requirement is satisfied for
$l \alt 10^8$  at the time $t_{\rm MR} \sim 6 \times
10^4~{\rm yr}$ of matter-radiation equality. In other words, by setting
$\tilde \lambda \sim 1$ and $m_l (t_{\rm MR}) \alt  1~{\rm MeV}$ we
find that the evolution of $m_l$ with cosmic time given in (\ref{mt}) is such
that at the last scattering surface the dominant KK state in the
dynamical dark matter ensemble has the correct decay width to
accommodate the CMB constraints. This is shown in Fig.~\ref{fig:1}
where we
display $\Gamma^l_{\gamma\gamma} (t)$ normalized to $m_l (t_{\rm MR})= 1~{\rm MeV}$. 

As can be seen in Fig.~\ref{fig:1}, for the selected set of fiducial parameters,
DDG predictions from KK decay into SM
fields saturate current upper limits placed by EDGES data on 
$\Gamma^l_{\gamma \gamma}$~\cite{Clark:2018ghm,Liu:2018uzy}. This
means that for this particular region of the parametr space, DDG predictions are within reach of next generation
experiments. In particular, interferometric
experiments like the upcoming Square Kilometre Array (SKA) will have
the sensitivity to make high-resolution spectra of high-redshift radio
sources to probe the 21-cm signal at Cosmic
Dawn~\cite{Weltman:2018zrl}. The photon emission from KK graviton decay could be distinguished from simple X-ray
heating by: {\it (i)}~occurring before the epoch of galaxy formation
and {\it (ii)}~by depositing energy more uniformly than would be
expected from galaxy clustering~\cite{Valdes:2007cu}. The lunar FARSIDE array will measure the Dark Ages global 21-cm
signal at redshifts $35< z <200$, extending the sensitivity down two orders of
magnitude below frequency bands accessible to ground-based radio
astronomy~\cite{Burns:2021pkx}. Thus, FARSIDE will provide an important test both of KK decays into SM fields and of the ideas discussed in this paper.

Once the free parameters of the model have been adjusted to
accommodate 21-cm data, we can use (\ref{Ggamma}) and (\ref{mt}) to
make predictions that can be confronted with observations in the local
universe. For example, for $m_l(t_{\rm MR}) \sim 1~{\rm MeV}$,
(\ref{mt}) leads to $m_l(t_{\rm today}) \sim 50~{\rm keV}$. Now, we
have seen that dark matter decay gives the daughter particles a
velocity kick. Self-gravitating dark-matter halos that have a virial
velocity smaller than this velocity kick may be disrupted by these
particle decays. Combined cosmological
zoom-in simulations of decaying dark matter with a  model of the Milky
Way satellite population rule out
non-relativistic kick speeds $\agt 10^{-4}$ for a dark matter
lifetime $\tau_{\rm DM} \alt 
29~{\rm Gyr}$ at 95\%CL~\cite{DES:2022doi}. However,  N-body simulations of
isolated dark-matter halos seem to indicate that if $\tau_{\rm DM} \agt
60~{\rm Gyr}$ and the kick
speed $\alt 10^{-2}$ then the halos are essentially
unchanged~\cite{Peter:2010jy}. Setting $\delta n \sim 1$~\cite{Cumrun} and taking $\beta
\sim 635$ to match our normalization ($l \sim 10^8$ at
$t_{\rm MR}$; see (\ref{mt})), we obtain  $\tau_{_{m_l (\rm today)}} \sim 68~{\rm Gyr}$
and so the kick velocity of the decay products, $v_{\rm r.m.s.}
\sim 5 \times 10^{-3}$, is consistent with the mass concentration of
self-gravitating dark matter halos.\footnote{Since the model is a
  realization of dynamical
  dark matter~\cite{Dienes:2011ja}, a dedicated simulation should be done to ensure full 
  compatibility with the evolution of structure formation.}

\subsection{Indirect Dark Matter Searches on Dwarf Spheroidals
and Cluster of Galaxies}

In line with our stated plan, we now adopt the parameter space
within reach of 21-cm data as benchmark and  we confront
  predictions of the DDG model with null results of indirect dark matter
  searches targeting
  dwarf spheroidal and cluster of galaxies. These
astrophysical environments are the best promising targets to search for
signals of decaying dark matter in the local universe, because they have a low content of
gas and dust, as well as a high mass-to-light ratio, and they are free
of astrophysical photon-emitting-sources. Besides, dwarf spheroidals are also
relatively nearby and many lie far enough from the Galactic plane to
have low Galactic foreground.

There are restrictive bounds on
$\Gamma^l_{\gamma \gamma}$ from null results of
indirect dark matter searches targeting dwarf spheroidals and cluster of
galaxies. In particular, null results from searches by the NuSTAR
Collaboration in the direction of the bullet cluster can be translated
into an upper limit $\Gamma^{\rm CDM}_{\gamma \gamma} < 
10^{-28}~{\rm s}^{-1}$ for dark matter masses in the 10 to 50 keV
range~\cite{Riemer-Sorensen:2015kqa}.

Analyzing the optimal region of the phase space explored in the
previous section, we find that for our choice of parameters, DDG
predictions saturate the bullet cluster  bound (see Fig.~\ref{fig:1}), suggesting that indirect detection dark matter experiments
are extremely competitive to detect the KK decay products in the local universe.

\subsection{Recapitulation}

We presented a two-step test of the DDG model. We have shown that for
$\tilde \lambda \sim 1$, $\delta n \sim 1$, $\beta \sim 635$, and
$m_l(t_{\rm MR}) \sim
1~{\rm MeV}$ the cosmic evolution of the  dynamical KK ensemble 
predicts via (\ref{mt}) a dominant particle mass of $\sim 900~{\rm
  keV}$ at CMB, of $\sim 500~{\rm keV}$ in the Dark Ages, of $\sim
150~{\rm keV}$ at Cosmic Dawn, and of $\sim 50~{\rm keV}$
in the local universe. This is in sharp contrast to typical dark matter decay scenarios with one unstable particle (such as sterile neutrinos~\cite{Abazajian:2017tcc}). For our fiducial parameters, the corresponding
decay widths of ${\rm KK} \to \gamma \gamma$ will be within reach of next generation experiments probing the low- and
high-redshift universe. Thus, a combination of positive results in
future observational campaigns can be used to pinpoint the DDG model.

\section{Higuchi bound and the shape of Inflation}
\label{sec:3}

In de Sitter (dS) spacetime (of radius $1/H$)  there exists an absolute
minimum for a field of spin-$2$ and mass $m_2$ set by the Higuchi bound
\begin{equation}
  m_2^2 \geq  2  H^2 \,,
\end{equation}
  where $H = \Lambda^2/M_{\rm Pl} \sim 10^{-34}~{\rm eV}$ is the Hubble parameter~\cite{Higuchi:1986py}. If
  the bound is violated the massive spin-2 field contains helicity
  modes with negative norm, which are in conflict with
  unitarity. Hence, if we define the ``size'' of the extra dimensions
  by the inverse mass of the lightest KK excitation of the graviton,
  the Higuchi bound forbids any compactification in which the extra
  dimensions are larger than and ${\cal O}(1)$ factor times
  $1/H$~\cite{Kleban:2015daa}.

Now, since $H \ll m_{\rm KK} \sim \Lambda^{1/4}/\lambda$, the Higuchi
bound is inordinately satisfied by the DDG construct today. However, 
the Higuchi bound forbids the presence of the DDG tower over a dS
background within the mass range $0 \leq m_l^2 \leq 2 H_I^2$, which would
imply that KK excitations could not be excited during
inflation. Indeed, the inflation scale is given by $M_I = \Lambda_I^{1/4}
= 3^{1/4} \sqrt{M_{*}^I H_I}$, where
$M_{*}^I = M_{\rm Pl}/\sqrt{N_I}$ is the strength of
gravity at the inflation scale, with
$N_I =M_*^I/m_{\rm KK}$ the number of species with masses below the inflation
scale. Now, using the standard formulae from slow roll single field inflation and
input from the experiment for the amplitude of density perturbations, the Hubble
factor during inflation is found to be $H_I
\sim 10^{-4} \ \sqrt{r} \ M_{*}^I$, with $r$ the tensor to
scalar ratio of primordial gravitational
waves~\cite{Antoniadis:2014xva}.\footnote{The latest CMB observations
  of BICEP/Keck, combined with those of WMAP and the Planck mission,
  place a strong upper bound $r \leq 0.036$ (at 95\%
  CL)\cite{BICEP:2021xfz}.} If the radion is fixed during inflation
(in 4 dimensions) the Higuchi bound gives $H_I \alt m_{\rm KK} \alt {\rm eV}$,
implying $M_I \alt 100~{\rm GeV}$. To confront this obstacle we adopt the working assumption that the Universe undergoes a period of inflation in which the radius of the dark
  dimension expanded exponentially fast,  from the species length up to the micron-scale.\footnote{Other aspects of inflation and higher spin states in relation with the Higuchi bound were discussed in~\cite{Scalisi:2019gfv}.}

The core idea behind the inflationary phase takes after the procedure
adopted in string theories with large internal dimensions and
TeV-scale
gravity~\cite{Antoniadis:1998ig,Arkani-Hamed:1999fet,Cline:1999ky}.
The higher-dimensional metric is given by
\begin{equation}
  ds_5^2 =  a_5^2 \ (-d\eta^2 + d\vec x^2 + r^2_0 \ dy^2) \,,
\label{metric}
\end{equation}
where $\eta$ is the conformal time, $a_5 = -1/(H\eta)$, $\vec x$ denotes the 3 uncompactified
dimensions, and $r_0 \sim 1/M_{\rm UV}$ is the radius of the
dark dimension $y$ at the beginning of the inflationary phase. The
4-dimensional decomposition in the Einstein frame is
found to be
\begin{equation}
  ds_5^2 = \frac{1}{R} ds_4^2 + R^2 dy^2 \,,
\label{metric2}  
\end{equation}  
where $ds_4^2 = a_4^2 (-d\eta^2 + d^2\vec x)$. Comparing
(\ref{metric}) and (\ref{metric2}) we arrive at $a_4/\sqrt{R} = R$. After inflation of $N$ $e$-folds,
where the scale factor was expanded by $a_5 = e^N$, the radius becomes
$R=e^N$. We want
$r_0$ to grow fast up to the micron scale. This requires 42
$e$-folds. Now, if $R$ expands $N$ $e$-folds, then the
4-dimensional space would expand $3N/2$ $e$-folds as a result of a uniform
5-dimensional inflation. Hence, 42 $e$-folds in $R$ implies 63 $e$-folds
in the non-compact space to successfully address the horizon
problem~\cite{Lidsey:1995np,ParticleDataGroup:2022pth}. We can now obtain an upper limit on the scale of inflation by noting that for $m_{\rm
   KK} \sim 1~{\rm eV}$, we have $M_*^I \sim 2 \times 10^9~{\rm GeV}$, and so
 $M_I = 10^{-2} M_*^I r^{1/4} \alt 8 \times 10^6~{\rm GeV}$.

The inflaton mass $m_\phi$ is model dependent, and we will take it as
a free parameter. To produce
matter on the brane it is sufficient to introduce a Yukawa-like
coupling $y$ of the inflaton to brane fermions, and so the decay width
is
\begin{equation}
  \Gamma^\phi_{f\bar f} \sim y^2 \ m_\phi \ \frac{m_{\rm KK}}{M_{\rm UV}} \,,
\label{GSM}
\end{equation}
where the last factor $m_{\rm
  KK}/M_{\rm UV}$ comes from the volume suppression. This suppression
is similar to the one in (\ref{in3}) for
the case of bulk right-handed neutrinos. The decay width into
gravitons is Planck suppressed and we have to carry out only one sum
up to $l$ because $\delta n \sim 1$. By dimensional analysis,
\begin{equation}
\Gamma^\phi_{\rm grav} \sim  \frac{m_\phi^3}{M_{\rm Pl}^2} \ \times \
\frac{m_\phi}{m_{\rm KK}} \ \times \ \delta n \sim
\frac{m_\phi^4}{M_{\rm UV}^3} \, .
\label{Ggrav}
\end{equation}
Comparing the decay rate of SM fields (\ref{GSM}) to that of gravitons
(\ref{Ggrav}) we impose 
\begin{equation}
  \left(\Gamma_{f \bar f}^\phi \sim \frac{m_{\rm KK}}{M_{\rm UV}} m_\phi \right)
> \left(\Gamma_{\rm grav}^\phi \sim 
  \frac{m_\phi^4}{M_{\rm UV}^3} \right) \,,
\label{comparison}
\end{equation}
where we have taken $y \sim 1$. All in all, (\ref{comparison}) gives a bound
for the mass of the inflaton $m_\phi < 1~{\rm TeV}$. The suppressed decay into gravitons ensures that the upper limit on
the number of ``equivalent'' light
neutrino species ($\Delta N_{\rm eff} < 0.214$ at 95\% CL~\cite{Planck:2018vyg}) present in the era before recombination is
satisfied~\cite{Anchordoqui:2020djl}.
The KK decomposition of the 5-dimensional metric gives rise to a tower
of massive physical
states of spin-2, while the radion is reduced only to the
4-dimensional zero mode. This is to be contrasted with the TeV-scale
model advocated in~\cite{Arkani-Hamed:1998sfv}, in which the inflaton
is a field localized on the brane. However, in both cases the inflaton decays re-heat predominantly brane-states while not producing significant numbers of gravitons.

\section{Conclusions}

\label{sec:4}

In the first part of this paper we generalized studies of large-extra-dimension
  models originally set up to interpret the smallness of neutrino
  masses by postulating that right-handed neutrinos, unlike SM fields,
  do propagate in the
  bulk~\cite{Dienes:1998sb,Arkani-Hamed:1998wuz,Dvali:1999cn,Davoudiasl:2002fq,Antoniadis:2002qm}. We
  showed that when interpreted in the
  context of the dark dimension, neutrino oscillation data constrain
  the characteristic mass scale of the KK tower: $m_{\rm KK} >
  2.5~{\rm eV}$ at the 99\%~CL. This constraint represents a two order of
  magnitude improvement over the bound on short-range deviations from
  Newton's gravitational law~\cite{Kapner:2006si,Lee:2020zjt}.

In the second part of the paper we proposed a method to test a
particular realization of the incredible bulk of dynamical dark
matter~\cite{Dienes:2011ja} in wich the ensemble of KK modes originates in a compact (dark)
dimension with characteristic length-scale in the micron range~\cite{Montero:2022prj,Gonzalo:2022jac}. Since
the cosmic evolution of the dark sector is driven by dark-to-dark
intra-ensemble decays, the model is particularly challenging to
probe. We developed a two-step process test to unambiguously
distinguish predictions of the dark dimension gravitons from those of the
$\Lambda$CDM model.  Measurements of redshifted 21-cm lines allow us to
perform a tomographic study of the Dark Ages and Cosmic Dawn to search
for the photon emission from KK graviton decay. We
showed that the changes induced by ${\rm KK} \to \gamma \gamma$ on the
21-cm signal could be within the reach of next generation experiments and
can be used to make the model fully predictable. In particular, we can
estimate the evolution of the dark graviton mass over the age of the
Universe to pin down its value today. We demonstrated that the
sensitivity of indirect dark matter searches targeting dwarf
spheroidals and cluster of galaxies are extremely competitive to
detect the KK decay products in the local universe.

In the last part of this paper we examined the global structure of the inflationary phase within the context of
  the dark dimension. We have shown that
the model parameters required for a successful uniform inflation driven by a 5-dimensional
  cosmological constant  (corresponding to a flat region of the
  5-dimensional potential) are natural.

In closing, we comment on how to make use of the DDG model
to try resolving an emerging tension between high- and low-redshift observations. The growth of cosmic structure is parametrized by
\begin{equation}
  S_8 = \sigma_8 \, \left(\frac{\Omega_m}{0.3}\right)^{0.5} \,,
  \label{S8}
\end{equation}
where $\sigma_8$ measures the r.m.s.
amplitude of the clustering of matter; more accurately,
\begin{equation}
  \sigma_8^2 = \frac{1}{2\pi^2} \int \frac{dk}{k} \ W^2(kR) \ k^3 \ P(k)\,,
\end{equation}
where $P(k)$ is the linear matter power spectrum at $z=0$ and $W(kR)$
is a top-hat filter describing a sphere (in Fourier space) with a
(historically chosen~\cite{Evrard:1989}) radius $R = 8~{\rm
  Mpc}/h$, and where $h$ is the dimensionless Hubble constant. Over the past few years, a tension has emerged between
$S_8$ as inferred by constraining $\Lambda$CDM parameters with CMB data ($S_8 = 0.834 \pm
0.017$)~\cite{Planck:2018vyg} and as measured from late-Universe
datasets; e.g., KIDS weak lensing surveys ($S_8 = 0.759 \pm
0.024$)~\cite{KiDS:2020suj} and DES combined analysis of the clustering of foreground galaxies and
lensing of background galaxies ($S_8 = 0.776 \pm
0.017$)~\cite{DES:2021wwk}.

It was first proposed in~\cite{Enqvist:2015ara} that the $S_8$ tension
can be relaxed if a fraction of CDM is unstable and decays into invisible massless
particles. These invisible particles, which interact
only via gravity with the visible SM sector, are generally referred to
as dark radiation (DR). It is evident that a reduction of $\Omega_m$
in (\ref{S8}) can help
accommodate the observed discrepancy in $S_8$.
However, as the decay into DR increases depleting the abundance of CDM at late times, the gravitational
lensing effect due to the evolving large-scale structure is
reduced. To accommodate the $S_8$ mismatch a high decay rate of ${\rm
  CDM} \to {\rm DR}$ is required and therefore the lensing effect is
markedly suppressed. This in turn implies that there is less transfer of
power on the CMB spectra with respect to the reference
$\Lambda$CDM. The decay ${\rm CDM} \to {\rm DR}$ is well-constrained by high-$\ell$ CMB
data because of the reduced small-scale anisotropies from the massless decay
products. In particular, data from the Planck mission lead to an upper
bound on the decay rate, 
$\Gamma^{\rm CDM}_{\rm DR}  < 10^{-19}~{\rm s}^{-1}
$ at 95\% CL~\cite{FrancoAbellan:2021sxk,Alvi:2022aam}, which is
insufficient to accommodate the $S_8$ tension.

Resolving the $S_8$ tension requires to decrease
the amplitude  of matter fluctuations on scales $k \sim  0.5 h/{\rm
  Mpc}$~\cite{FrancoAbellan:2020xnr}. This can be achieved with decaying CDM into one
massive warm dark matter (WDM) particle and one massless particle, both
interacting only through gravitation with SM fields. A critical
difference with the previous scenario is that the massive daughter particle could be born
relativistic at $z_{\rm decay}$, when the expansion rate is given by
$H(z_{\rm decay})$, but behave like CDM as the universe evolves. The relativistic massive decay product would suppress the formation of cosmic
structure, because at production it free-stream with finite kick-velocity and can escape from the gravitational potential wells surrounding matter over-densities. The decay of ${\rm CDM} \to {\rm DR} + {\rm WDM}$ produces a change in
both $\sigma_8$ and $\Omega_m$ and can accommodate the $S_8$ tension if
$\Gamma^{\rm CDM}_{\rm DR + WDM} \sim 6 \times 10^{-19}~{\rm s}^{-1}$~\cite{FrancoAbellan:2020xnr}. This is
because the WDM component partially contributes to the
matter energy density slowing down the lensing suppression of the
CMB~\cite{FrancoAbellan:2021sxk}. Note that such a decay width  corresponds to a lifetime
$\tau_{\rm DM} \sim 53~{\rm Gyr}$. More recent analyses suggest that the
parameter space that can accommodate the $S_8$ tension favored by
current data points to $\tau_{\rm DM} \sim
100~{\rm Gyr}$ and $v_{\rm r.m.s.} \sim
10^{-3}$~\cite{Simon:2022ftd,Fuss:2022zyt,Tanimura:2023bkh}. While
the fiducial values entertained in the previous section ($\tau_{m_l(\rm today)}
\sim 68~{\rm Gyr}$ and $v_{\rm r.m.s.} \sim 5 \times 10^{-3}$) are in the ballpark and
it is tempting to investigate whether the DDG model can resolve the $S_8$, we
note that any scan of the DDG parameter space to try to accomodate the
$S_8$ requirements would be in need of a full-scale numerical simulation considering the evolution of
the intra-ensemble KK decays  
throughout the history of the universe. Such an interesting task is beyond the scope
of the present study.

\section*{Acknowledgments}

We have greatly benefited from discussions with Nima Arkani-Hamed, Miguel Montero and
Cumrun Vafa. The work of L.A.A. is supported by the U.S. National
Science Foundation (NSF Grant PHY-2112527). The work of D.L. is
supported by the Origins Excellence Cluster and by the
German-Israel-Project (DIP) on Holography and the Swampland.

\end{document}